\documentstyle[epsfig,12pt]{article}
\newlength{\dinwidth}
\newlength{\dinmargin}
\def\beq{\begin{equation}}
\def\eeq{\end{equation}}
\def\bea{\begin{eqnarray}}
\def\eea{\end{eqnarray}}
\def\barr{\begin{array}}
\def\earr{\end{array}}
\def\d{\delta}
\def\m{\mu}
\def\n{\nu}
\def\a{\alpha}
\def\b{\beta}

\newcommand{\signal}{\mbox{$e^-_L e^-_L \to W^- W^-$}}
\newcommand{\revsignal}{\mbox{$e^+_R e^+_R \to W^+ W^+$}}

\begin{document}
\thispagestyle{empty}
\addtocounter{page}{-1}
\begin{flushright}
DESY 96-253\\
BUTP-96/28\\
hep-ph/9612340\\
December 1996
\end{flushright}
\vspace*{1.8cm}
\centerline{\Large\bf
Heavy Majorana neutrinos in $e^-e^-$ collisions
\footnote{Work supported in part by Schweizerischer
Nationalfonds}}
\vspace*{2.0cm}
\centerline{\large\bf Christoph Greub }
\vspace*{0.4cm}
\centerline{\large\it Deutsches Elektronen-Synchrotron DESY, Hamburg}
\vspace*{0.8cm}
\centerline{\large\bf Peter Minkowski}
\vspace*{0.4cm}
\centerline{\large\it Institute for Theoretical Physics, University of 
Bern,}
\centerline{\large\it Sidlerstr. 5, CH-3012 Bern, Switzerland}
\vspace*{2.5cm}
\centerline{\Large\bf Abstract}
\vspace*{1cm}
We discuss the process 
$e^-e^- \to W^- W^-$ mediated by heavy 
Majorana neutrino exchange in the $t-$ and $u$ channel.
In our model the cross section for this reaction is a function of  
the masses ($m_N$) of the heavy Majorana neutrinos and 
mixing parameters ($U_{eN}$) originating from mixing
between the ordinary left-handed standard model neutrinos
and additional singlet right-handed neutrino fields.
Taking into account 
the standard model background and contraints from low energy
measurements, we present discovery limits in the ($m_N,U_{eN}^2$) plane.
We also discuss how to measure in principle the 
CP violating phases, i.e., the relative phases
between the mixing parameters. 

\vspace*{1.5cm}
\centerline{(To be published in the Proceedings of the 1996
DPF/DPB Summer Study}
\centerline{ on New Directions for High Energy
Physics (Snowmass 96).)}

\newpage

\section{Introduction}
The question why the masses of the observed 
neutrinos are much lighter than those
of the charged leptons 
is a central unsolved problem in particle physics.
An attractive scenarios to understand this is the following:
Adding 
right-handed singlet neutrino fields  to the standard
model, the resulting mass spectrum 
can be such that there are  3 essentially zero mass Majorana neutrinos
and additional heavy Majorana neutrinos. While in the generic case
the masses of these heavy neutrinos are too large
to detect observable consequences at present and future colliders,
it not excluded however, that
their masses are in the range of a few TeV
and that their coupling strengh to the charged leptons are rather
large. For the latter case, the phenomenology for
$e^+e^-$ collisions where 
heavy Majorana neutrinos can be directly produced 
via $t$- channel $W$  and $s$- channel $Z^0$ exchange,
has been worked
out in detail in \cite{buchmueller}. The main conclusion is
that the cross sections and designed luminosities 
are large enough to find heavy Majorana neutrinos essentially
up to the kinematical limit, i.e., up to $m_N \approx \sqrt{s}$,
provided that the relevant mixing angles are near the present 
bounds inferred from low energy data. A similar conclusion
also holds for single Majorana neutrino production in $e \gamma$
collisions. Another attractive reaction to be discussed in
detail in this paper is $e^-e^- \to W^- W^-$, which is mediated
by $t$- and $u$- channel Majorana neutrino exchange (see Fig. 1).
As the neutrinos do not have to be produced in this reaction,
one is potentally sensitive to neutrino masses which exceed 
$\sqrt{s}$. Therefore, the aim of this paper is to work out 
the discovery region for such neutrinos in the parameter space 
(masses and mixing angles), taking into account standard model
background reactions. 

In the second part of this introduction we present the model
and list the present bounds on masses and mixing angles inferred
from low energy precision measurements.
In section II we discuss the signal reaction $e^-e^- \to W^-W^-$,
while section III deals with the simulation of the standard model
background. In section IV we present the discovery limits
in the $(m_N,U_{eN}^2)$ plane.
Section V finally deals with the question of how one can measure
in principle
the relative phases of the mixing angles; the presence of such phases
is a necessary condition to observe CP violation in reactions like
$e^-e^- \to W^-W^-$.

\subsection{The Model}
As in the standard model (SM) only left-handed neutrino fields
are present, a Dirac mass $(m_D)$ term for the neutrinos
cannot be written down;
also a Majorana mass term $(m_M)$ involving only the 
left-handed neutrino fields ($\nu_L$) cannot be constructed consistently
within the SM because this would require
Higgs-Triplets which are absent in the SM.
Consequently, neutrinos are exactly 
massless in the SM.
There are, however, many extensions of the SM (like $SO(10), E_6,...$)
with $k$ extra
neutrino fields ($\nu_R$) which are singlets 
under the SM gauge group $G_{SM}$.
In a generic case symmetry breaking then induces the mass term
\beq
\label{massterm}
L_\nu^{mass} = - \overline{\nu_L} m_D \nu_R - \frac{1}{2}
\overline{\nu_R^c} \, m_M \, \nu_R + h.c \quad ,
\eeq 
where $m_D$ is a $(3 \times k)$ Dirac mass matrix and
$m_M$ is a $(k \times k)$ Majorana mass matrix. 
As an example, the grand unified group SO(10)
can break to the SM group through the chain
\beq
SO(10) \stackrel{\Lambda_{GUT}}{\rightarrow}
  G_{SM} \times U(1)_{B-L} \stackrel{v'}{\rightarrow} G_{SM}
\stackrel{v}{\rightarrow} SU(3)_c \times U(1)_{em} \quad ,
\eeq
where $\Lambda_{GUT}$, $v'$ and $v=174$ GeV are the respective
breaking scales in this chain.
In this example, the Dirac- and Majorana mass matrices
are of the form 
\beq
\label{massform}
m_D = g \cdot v \quad , \quad m_M = h \cdot v' \quad ,
\eeq 
where $g$ and $h$ are matrices of Yukawa couplings.
It is conceivable that the residual $B-L$ breaking scale $v'$ is as low as
a few
$TeV$; in this case also the Majorana mass term is expected to
be relatively light. 

To get the mass eigenstates, one has to diagonalize
the neutrino mass matrix by means of a unitary ($3+k \times 3+k$)
matrix $U$. The resulting mass eigenstates are 3 light Majorana
neutrinos $\nu_i$ with masses $m_{\nu_i}$ and $k$ heavy
Majorana neutrinos $N_i$ with masses $m_{N_i}$.
Expressing the weak eigenstates in the charged current
by the corresponding
mass eigenstates \footnote{We assume without loss of generality
that the mass matrix of the charged leptons
has been diagonalized prior to the neutrino mass matrix.}, 
we get the $W-e-N$ ($W-e-\nu$) vertices
relevant for our process $e^-e^- \to W^- W^-$:
\beq
\label{vertex}
L_{We\nu,WeN} = \frac{g}{\sqrt{2}} \, 
\left( \sum_\nu \bar{e} \gamma_\mu L U_{e\nu} \nu +
\sum_N \bar{e} \gamma_\mu L U_{eN} N \, \right) \, W^\mu \quad .
\eeq    
It is instructive to look for a moment at the simpler  $1+1$
dimensional case
with only one left-handed and one right-handed neutrino field.
The masses and the mixing angle $U_{eN}$ are given by
the see-saw formulae \cite{seesaw}
\beq
\label{massmix}
m_\nu \approx \frac{m_D^2}{m_M} \quad , \quad
m_N \approx m_M \quad , \quad
U_{eN} \approx \frac{m_D}{m_M} \quad . 
\eeq 
In this case, the bounds on the light neutrinos imply that
the mixing angle $U_{eN}$ has to be very small; consequently
the cross section in $e^-e^- \to W^- W^-$ is much too small
to be observed. 
In the general $3+k$ dimensional case, however, it is possible,
that the 3 light neutrinos are exactly massless 
while
the mixing angles $U_{eN}$ remain finite (i.e. nonzero)!
This means that there is some chance to find heavy neutrinos
with rather large cross sections
through the process $e^-e^- \to W^- W^-$.

In the following, we assume the light neutrinos to be 
massless (minimal mass generating case) and the mixing angles
$U_{eN}$ and heavy masses $m_N$ to be essentially free parameters.   

\subsection{Constraints on masses and mixing angles}
Before listing the constraints from low energy measurements,
we would like to point out a model constraint.
Assuming that there are no Higgs triplet fields, the 
Majorana mass term for the left-handed
neutrinos is absent (as in eq. (\ref{massterm})). This
implies the relation
\beq
\label{modconstraint}
\sum_\nu m_\nu U_{e \nu}^2 +  \sum_N m_N U_{eN}^2 = 0 \quad , 
\eeq
or in the minimal mass generating case ($m_\nu=0$)
\beq
\label{modmin}
 \sum_N m_N U_{eN}^2 = 0 \quad . 
\eeq

We now list the experimental contraints:
\begin{itemize}
\item
Charged current universality implies  
that
$|U_{eN}|^2 < 4 \times 10^{-3}$.
The LEP-1 data lead to similar bounds \cite{Rizzo}.  
\item
Heavy Majorana neutrinos mediate the rare decay $\mu \to e \gamma$.
Non-observation of this decay leads to the strong bound  
$\left|\sum_N U_{\mu N} U^*_{eN} \right| < 2 \times 10^{-4}$. 
This bound, however, does not necessarily lead to a more stringent
bound on $U_{eN}$ than charged current universality.
\item
The non-observation of neutrinoless double beta decay
$\beta \beta_{0\nu}$ implies a bound on both, the light and heavy
neutrino sector.
The best present evidence on the non-observation of the
$\beta \beta_{0\nu}$ reaction is from the experimental limit on the
process $^{76}Ge \to ^{76}Se e^-e^-$, with 
$\tau_{1/2} > 5 \times 10^{24}$ years.
For light neutrinos the bound reads
\beq
\label{betalight}
\left| \sum_\nu U_{e\nu}^2 m_\nu \right| \le \mbox{few eV} \quad.
\eeq 
In the minimal mass generating case ($m_\nu=0$) this bound is satisfied
automatically.
For heavy Majorana neutrinos the bound is of the form
\beq
\label{betaheavy}
\left| \sum_N \frac{U_{eN}^2}{m_N} \right|^2 \times
|\mbox{hadr. matrix element}|^2 \le \frac{1}{\tau_{1/2}} 
\quad.
\eeq 
A recent paper \cite{Minkheusch} shows that in the older 
literature the hadronic matrix elements have been strongly
overestimated. Taking the new estimates for the hadronic
matrix elements, the bound
becomes
\beq
\label{betaheavybound}
\left| \sum_N \frac{U_{eN}^2}{m_N} \right|  <
6 \times 10^{-3} \ TeV^{-1} 
\quad.
\eeq 
\end{itemize}  

\section{Signal}
In the minimal mass generating case, where the light neutrinos
are exactly massless by definition, 
only the heavy Majorana neutrinos $(N_i)$ contribute
to the process \signal. 
The corresponding Feynman diagrams are shown in Fig. 1.
\begin{figure}[htb]
\vspace{0.10in}
\centerline{
\epsfig{file=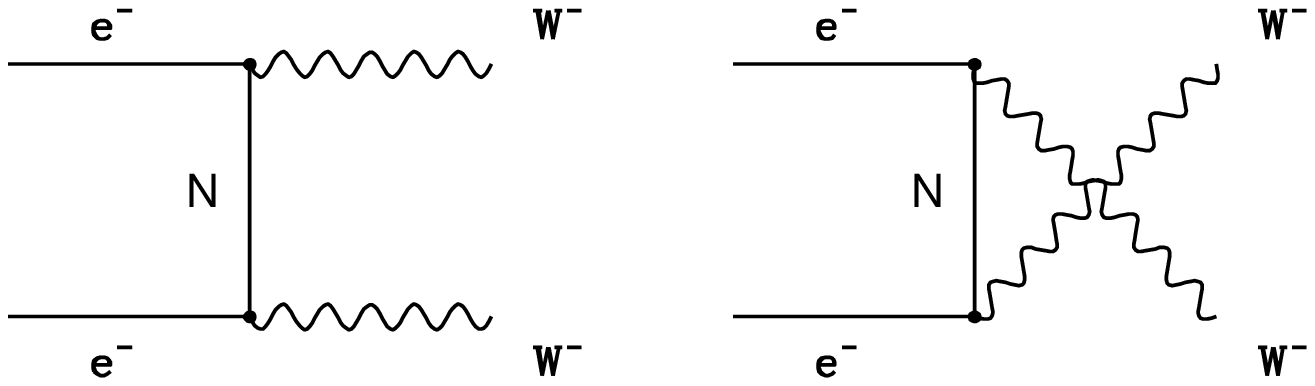,width=3.5in,angle=0,clip=}
}
\vspace{0.08in}
\caption[]{Feynman diagrams for the process \signal.}
\end{figure}
For $\sqrt{s} \gg m_W$ the cross section is totally 
dominated by the production of longitudinally polarized $W^-$.
To leading order in $m_W/\sqrt{s}$ and in tree approximation
the angular distribution
$d\sigma(e^-_L e^-_L \to W^- W^-)/d\cos \theta $ 
is given by the expression \cite{signal}
\beq
\label{production}
\frac{d\sigma}{d \cos \theta} = \frac{G_F^2}{8 \pi }
\, \left| \sum_{N} m_{N} \, U_{eN}^2 \, \left( \frac{t}{t-m_N^2} +
\frac{u}{u-m_N^2} \right) \right|^2 \quad ,
\eeq
with
$t=-(s/2) \, (1- \cos \theta)$ and $u=-(s/2) \, (1+ \cos \theta)$.
The general relation between mass and mixing parameters 
given in eq. (\ref{modconstraint}) 
for all neutrino flavors 
ensures unitarity and renormalizability. 
As a consequence of eq. (\ref{modmin}) [or eq. (\ref{modconstraint})], 
the cross section in eq. (\ref{production})  
vanishes in the limit 
$\sqrt{s} \to \infty$ 
because the leading constant term 
becomes proportional to $\sum_N m_N U_{eN}^2$. 
Another feature which is easily seen from 
eqs. (\ref{production}) and (\ref{modmin}) is 
that the heavy Majorana neutrinos have to be degenerate
in mass in order to have a non-zero cross section.

Note, that in the limit where the masses of all heavy Majorana 
neutrinos are much 
larger than $\sqrt{s}$, formula 
(\ref{production}) could be further approximated to read
\beq
\label{prodapp}
\frac{d\sigma}{d \cos \theta} = \frac{G_F^2}{8 \pi}
\, \left| \sum_{N}  \frac{U_{eN}^2}{m_N} \right|^2 \, s^2 \quad ,
\quad m_W \ll \sqrt{s} \ll m_N \quad .
\eeq
In the subsequent numerical analysis the lightest of the heavy neutrino
masses is singled out for
simplicity in the scenario
\beq
\label{scenario}
m_W \ll \sqrt{s} \le m_{N_1} \ll m_{N_2}, m_{N_3}, \ldots \quad .
\eeq
The results are then parametrized by the mass $m_N (=m_{N_1})$ and the 
corresponding mixing angle $U_{eN}$.

In the reaction \signal the invariant mass 
of the $W^-W^-$ pair equals $\sqrt{s}$; this is in contrast to 
all the
background reactions we discuss later, where the invariant mass
of the $W^-W^-$ pair 
reaches $\sqrt{s}$ only in very special kinematical configurations. 
Therefore,
this invariant mass  is a essential variable to get rid of the background.
To fully exploit this, we  concentrate on the hadronic decay modes of
the $W^-W^-$ boson pair where all the decay products can be detected (no 
neutrinos as in the purely leptonic or semileptonic decay mode).
The branching ratio of the $W^-W^-$ pair into the three possible 
channel is given roughly
by
\bea
\label{branching}
BR(W^-W^-: \mbox{both W decay hadronically}) &=& \frac{36}{81} \nonumber \\ 
BR(W^-W^-: \mbox{both W decay leptonically}) &=& \frac{9}{81} \nonumber \\ 
BR(W^-W^-: \mbox{one decays hadr., the other lept.}) 
&=& \frac{36}{81} \nonumber
\eea

Before discussing the background it is instructive to see that
the number of signal events (using eqs. (\ref{production}) and 
(\ref{branching})) 
can be rather large:
For $\sqrt{s}=500$ GeV, $m_N =1$ TeV and $U_{eN}^2=4 \times 10^{-3} $
the production cross section for $W^-W^-$ followed by hadronic decay
of both $W^-$ is $\sigma = 1.2$ fb; this corresponds to 60 events 
assuming a luminosity of $L=50 \, fb^{-1}$. 

\section{Background}
The most important backgrounds come from the  reactions 
$ e^- e^- \to W^- W^- \nu \nu$ (1),
$ e^- e^- \to W^- Z^0 \nu e^-$ (2), 
$ e^- e^- \to Z^0 Z^0 e^- e^-$ (3), and 
$ e^- e^- \to W^- W^+ e^- e^-$ (4),
where the gauge bosons decay hadronically and the
charged leptons escape along the beam pipe. If the jet charges
could be reconstructed, then of course only reaction (1)
would remain as a backgound. 
In this study we assume that 
this is not possible, hence deriving rather 
conservative discovery limits.
While (1), (2) and (3) have  been calculated 
by Cuypers et al. \cite{Cuypers}, the process (4) has been
included by Barger et al.  \cite{Barger}.
As mentioned earlier, the invariant mass of the two vector bosons
is in general much smaller than $\sqrt{s}$, because
of the additional leptons in the final state (see also Fig. 3).
However, there is a tail in the invariant mass distributions of the
gauge boson pairs in the background reactions  extending up to 
$\sqrt{s}$, because the additional leptons are essentially massless.
Given the fact that the  cross section of 
the background reaction (4) is more than
two orders of magnitude larger
than the signal (using optimistic
parameters for $m_N$ and $U_{eN}$)
one should try   
to further reduce the background by imposing additional suitable
cuts which do not affect the signal significantly.
To do such studies, we wrote a Montecarlo simulation \cite{Ali} 
of the
signal and the background. Reaction (4) is clearly the most serious
background source, because first, it has by far the largest
cross section \cite{Barger} and second, this cross section is dominated
by configurations where the final state electrons disappear
in the beampipe.
We therefore only take into account this background reaction.
We calculated reaction (4) in the Weizs\"acker-Williams approximation
for left-handed electron beams, which means that we have to 
work out in a first step the cross sections for the subprocess
$\gamma \gamma \to W^+ W^- \to jets$
for polarized on-shell photons.
By calculating directly the matrix elements with 4 fermions in the
final state and then doing the narrow width approximation of the
$W$-propagators we have correctly incorporated the polarization
effects of the intermediate vector bosons.
In a second step we 
convolute  the $\gamma \gamma$ cross sections
with the appropriate Weizs\"acker-Williams photon distribution functions.
The corresponding three Feynman diagrams are shown in Fig. 2.
\begin{figure}[htb]
\vspace{0.10in}
\centerline{
\epsfig{file=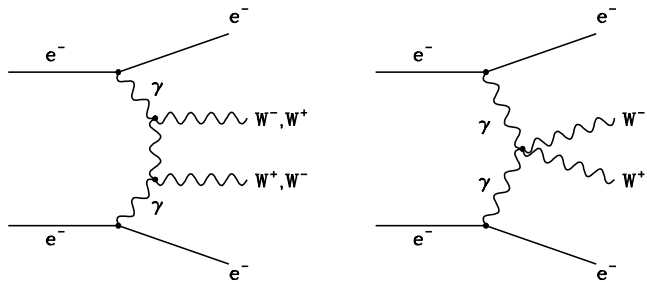,width=3.5in,angle=0,clip=}
}
\vspace{0.08in}
\caption[]{Feynman diagrams for the background process
$e^-_L e^-_L \to W^+ W^- e^- e^-$. Only those three diagrams are 
shown which contribute 
in the Weizs\"acker-Williams approximation.}
\end{figure}
As a check, we compared our Weizs\"acker-Williams results for 
$e^- e^- \to W^- W^+ e^- e^-$ with the exact results in ref.
\cite{Barger} and found good agreement. 
The resulting invariant mass distribution of the final state 
hadrons is shown in Fig. 3.
\begin{figure}[htb]
\vspace{0.10in}
\centerline{
\epsfig{file=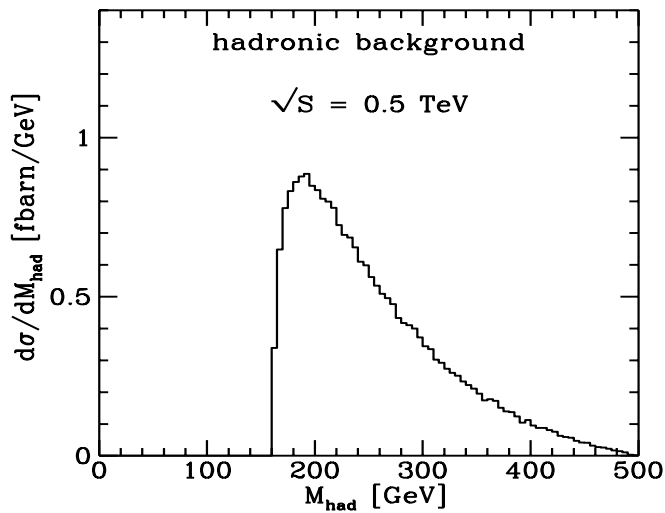,height=3in,width=3.5in,angle=0,clip=}
}
\vspace{0.08in}
\caption[]{Invariant hadronic mass distribution 
in the process $e^-_L e^-_L \to e^- e^- W^- W^+ \to e^- e^- 
+ jets$ in the Weizs\"acker-Williams approximation.}
\end{figure}
Needless to say, this reaction is only a  background, if the
final state electrons disappear in the beam direction. However,
as seen in Fig. 4 of ref.  \cite{minkhiggs} in these proceedings, 
this background can only be reduced  by a few
percent when imposing a
realistic angular cut on the electron directions. 

In order to reduce the background we require the hadronic mass
$m_{had}$ to be larger than a certain critical value $m_{had}^{crit}$. 
For each
$\sqrt{s}$ we do the analysis for the two typical values  
$m_{had}^{crit}=0.90 \times \sqrt{s}$ and  
$m_{had}^{crit}=0.80 \times \sqrt{s}$, respectively.
Figs. 4 and 5 illustrate that the typical average transverse momentum
$p_\perp(aver.)$
of the four jets, defined as $p_\perp(aver.) = 
(1/4) \, [|p_\perp|(1)+ \ldots + |p_\perp|(4)]$  
is higher for the signal  than for the background 
(after imposing the $m_{had}^{crit}$-cut).
  \begin{figure}[htb]
\vspace{0.10in}
\centerline{
\epsfig{file=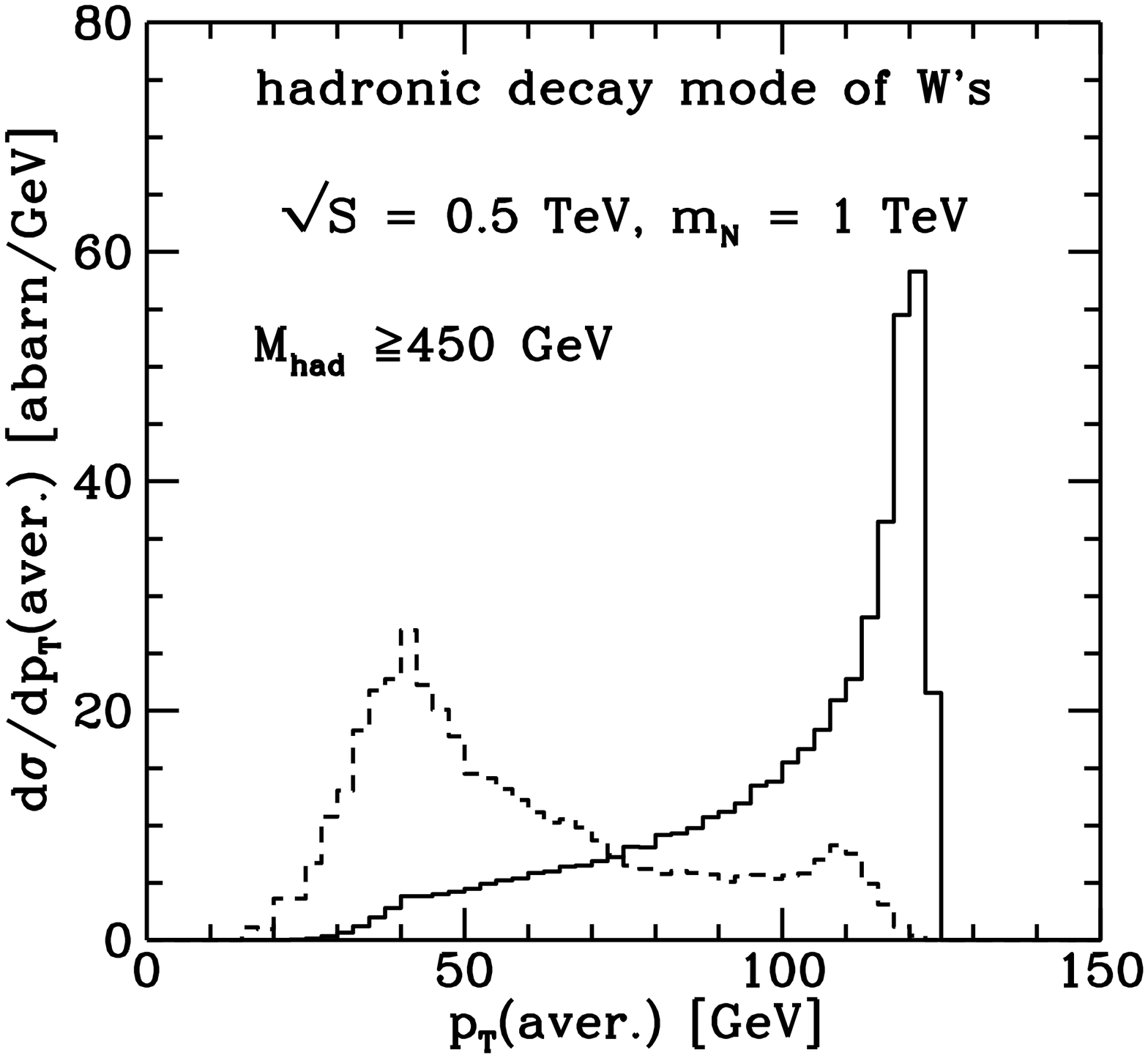,height=3in,width=3.5in,angle=0,clip=}
}
\vspace{0.08in}
\caption[]{Signal (solid line) and background (dashed line) 
as a function
of the average transverse momentum $p_\perp(aver.)$ of the jets.
The hadronic invariant mass is required to be $m_{had} \ge 450$ GeV
and the other parameters are chosen to be
$\sqrt{s}=500$ GeV, $m_N=1$ TeV, $U_{eN}^2 = 4 \times 10^{-3}$. 
}
\end{figure}

\begin{figure}[htb]
\vspace{0.10in}
\centerline{
\epsfig{file=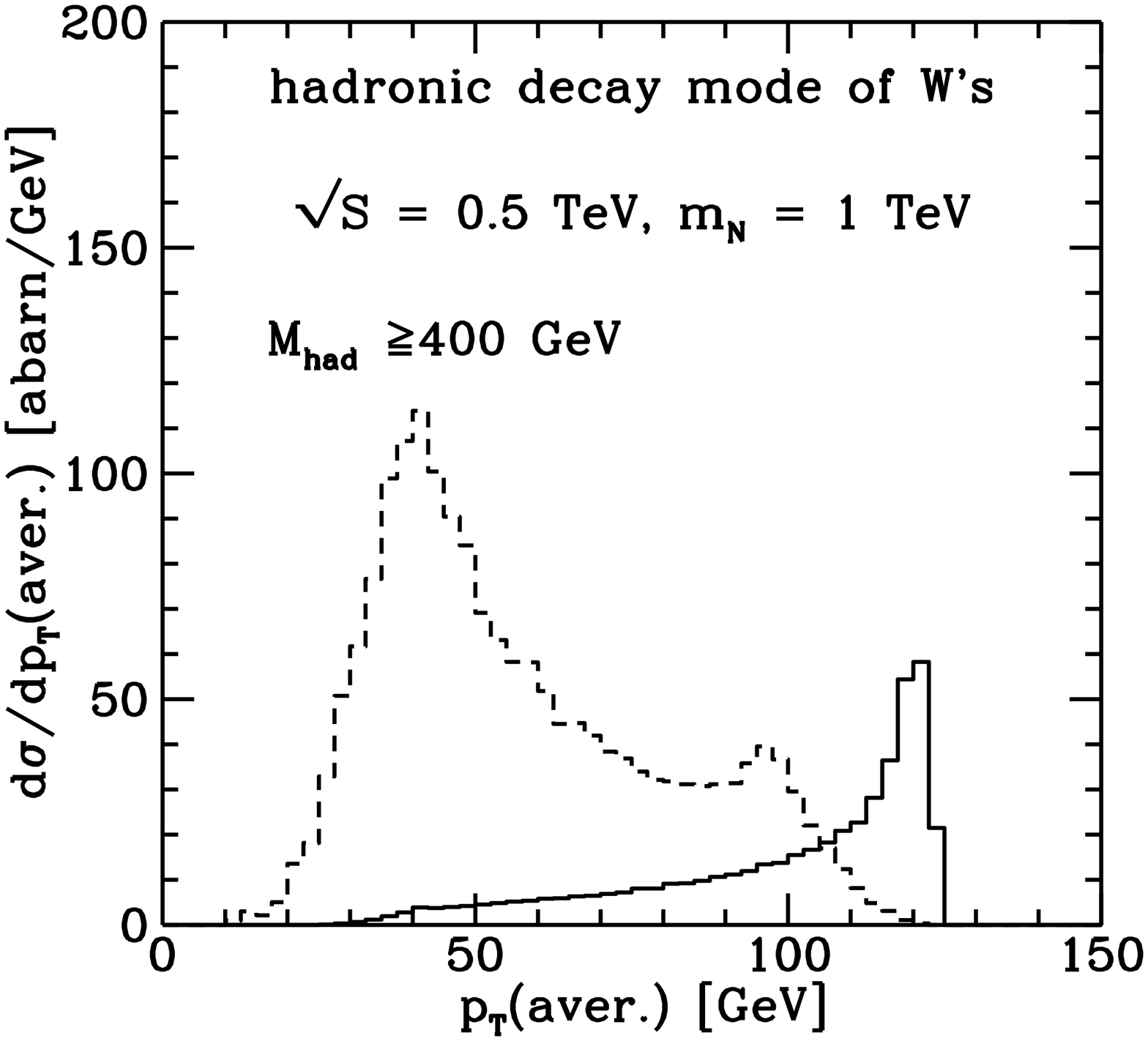,height=3in,width=3.5in,angle=0,clip=}
}
\vspace{0.08in}
\caption[]{Same as Fig. 4, but for $m_{had} \ge 400$ GeV.}
\end{figure}
One therefore can further enhance the signal/background ratio 
by imposing a lower cut on the variable $p_\perp(aver.)$.

\section{Discovery Limits}
To summarize, imposing cuts on the invariant hadronic mass, 
$m_{had}$, and
the average transverse momentum of the jets, $p_\perp(aver.)$, 
the background to
the signal process $e^-_L e^-_L \to W^- W^- 
\to \mbox{jets}$, which is characterized by 
the two parameters $m_N$ and $U_{eN}$, can be significantly
reduced.
As discussed in section I, there are constraints from low energy
data. While neutrinoless double beta decay
 is the most stingent constraint (see eq. 
(\ref{betaheavybound})) for neutrino masses below 660 GeV,
the bound from charged current universality 
($U_{eN}^2 < 4 \times 10^{-3}$)
dominates for larger masses. As in our plots the mass is mostly
larger than 660 GeV we did not plot the bound from 
$\beta \beta_{0\nu}$. 
Requiring a signal/backgound ratio $>1$ and the number
of signal events to be $>20$, we worked out the discovery limits
in the $(m_N,U_{eN}^2)$-plane.
\begin{figure}[htb]
\vspace{0.10in}
\centerline{
\epsfig{file=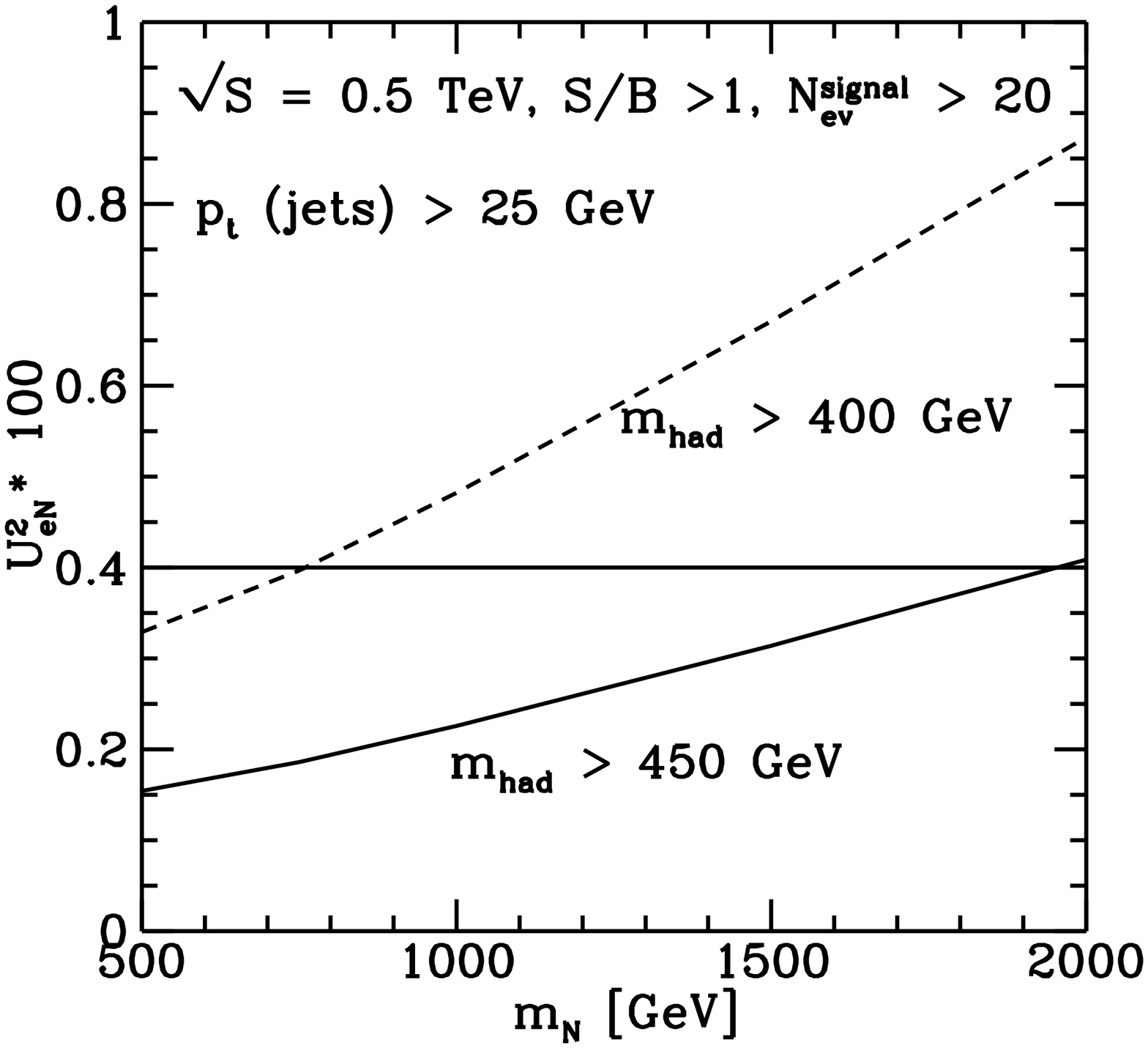,height=3in,width=3.5in,angle=0,clip=}
}
\vspace{0.08in}
\caption[]{Discovery limits in the $(m_N,U_{eN}^2)$-plane
for $\sqrt{s}=500$ GeV and luminosity
$L=50 \, fb^{-1}$. 
The horizontal solid line at $U_{eN}^2 = 4 \times
10^{-3}$ represents the upper bound on $U_{eN}^2$ 
from charged current universality.
Requiring $p_\perp(aver.)>25$ GeV and
$m_{had}>450$ GeV, the allowed region in the parameter space is above
the solid curve and below the horizonal line. Using the looser cuts 
$p_\perp(aver.)>25$ GeV and
$m_{had}>400$, the allowed region lies above the dashed
curve and below the horizontal line.}
\end{figure}
\begin{figure}[htb]
\vspace{0.10in}
\centerline{
\epsfig{file=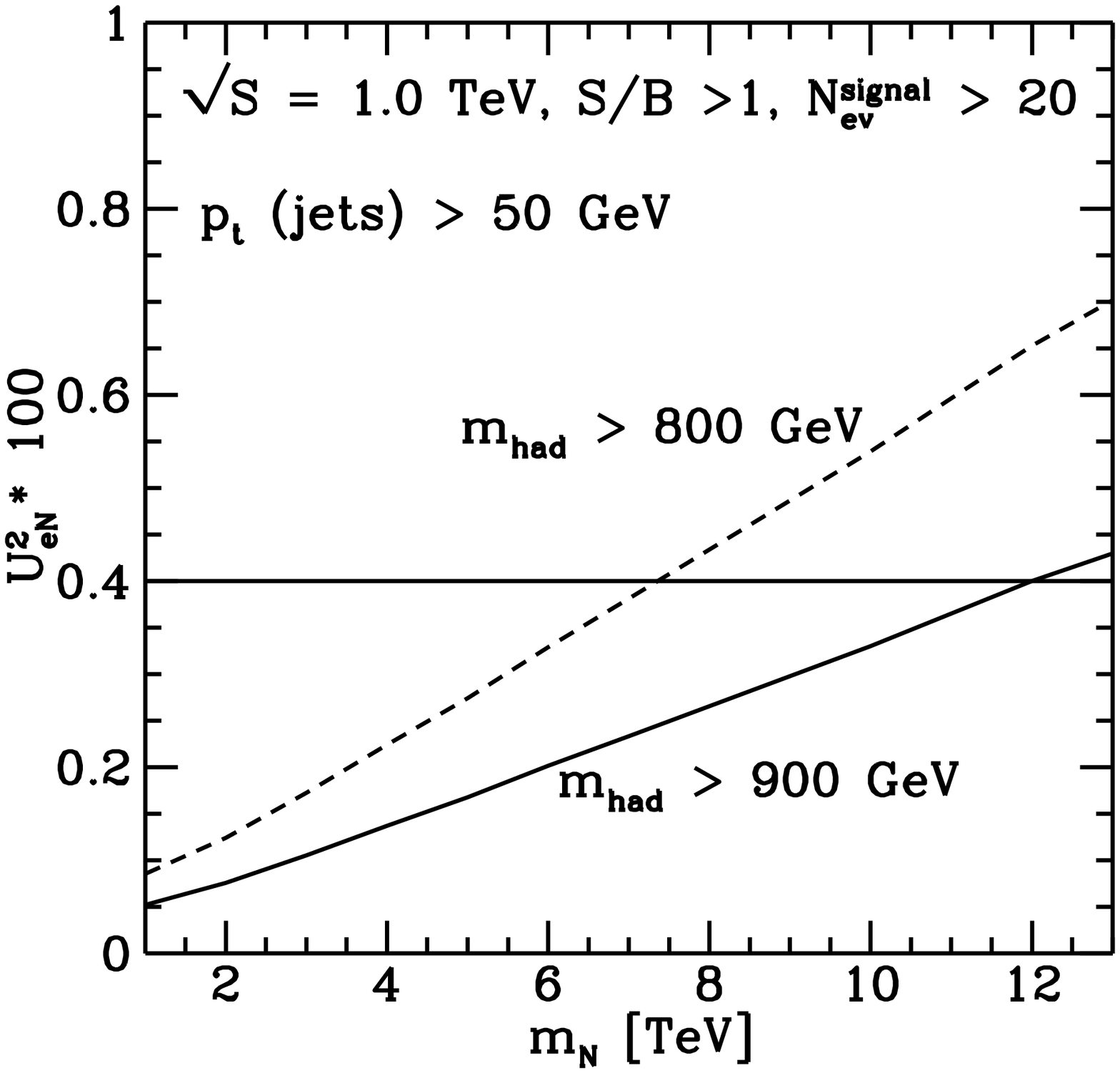,height=3in,width=3.5in,angle=0,clip=}
}
\vspace{0.08in}
\caption[]{As Fig. 6, but for $\sqrt{s}=1$ TeV and luminosity
$L=100 \, fb^{-1}$. The solid line corresponds to 
$m_{had} > 900 $ GeV and $p_\perp(aver.) > 50$ GeV, 
while the dashed line is for
$m_{had} > 800 $ GeV and $p_\perp(aver.) > 50$ GeV. The horizontal line
again represents the upper bound on $U_{eN}^2$ 
from charged current universality.}
\end{figure}
The results for $\sqrt{s}=500$ GeV are shown in Fig. 6,
assuming a luminosity $L=50 \,fb^{-1}$ and requiring the average
momentum of the jets $p_\perp(aver.)$ to be $> 25$ GeV.
In this figure the horizontal line represents the bound from
charged current universality and/or LEP-1 data. 
Imposing the cut on $m_{had}$ at 450 GeV the  allowed
region in the parameter space is above the solid curve
and below the horizontal line. If $m_{had}$ is cut at
400 GeV the allowed region lies between the dashed curve
and the horizontal line. From the figure one sees, that
it is very crucial at which value the cut on $m_{had}$ is imposed.
Cutting at 400 GeV, the allowed parameter space is essential
empty; if the cut is imposed at 450 GeV, one is sensitive
to a rather large parameter space: masses up to 2 TeV 
could be discovered. 
In an ideal experimental situation, without any fluctuations
in $\sqrt{s}$, one would of course make the $m_{had}$-cut
basically at $\sqrt{s}$. In the real situation
this energy is smeared, however, by beamstrahlung and initial
state radiation. Of course one could simulate these effects in order
to find out the optimal value at which
the $m_{had}$-cut should be imposed. Also the cut on the average
transverse momentum was imposed in a rather arbitrary, but 
reasonable way;
also this cut could be optimized in order to become sensitive
to a somewhat larger parameter space.
 
The corresponding results 
for $\sqrt{s}= 1$ TeV and $L=100 \, fb^{-1}$ are shown in Fig. 7. 
It is impressive that one is sensitive to neutrino masses up to
12 TeV, assuming the mixing angle at its present upper
bound.  

\section{CP violating phase from the interference region}
In this section we wish to discuss 
an alternative
scenario \cite{PM:sc}, \cite{PManant}, with {\bf two} relatively light
heavy neutrinos according to
\beq
\label{scenario2}
\label{6}
m_W \ll \sqrt{s} \stackrel{\le}{>} m_{N_1} 
\le m_{N_2} \ll m_{N_3}, \ldots \quad .
\eeq
In this case an interference can arise
at the level of the single helicity amplitude 
$e^-_L e^-_L \to W^-_L W^-_L$.
To make this more explicit,
we rewrite the cross section (\ref{production}) using the parametrization
\beq
\label{param2}
\label{7}
\begin{array}{l}
m_{N_1} \, U_{eN_1}^2 = e^{i \lambda}  
\left \lbrack \, a \, e^{i \delta /2} \ \right \rbrack
\ ; \ 
m_{N_2} \, U_{eN_2}^2 = e^{i \lambda} \left 
\lbrack \, b \, e^{-i \delta /2} \ \right \rbrack
\vspace*{0.2cm} \\
q_1 = \left ( m_{N_1} \right )^{2}
\ ; \ q_2 = \left ( m_{N_2} \right )^{2}
\ ; \ q_1 < q_2  
\ , \ a , b \ge 0 \ (\mbox{real})
\quad .
\end{array}
\eeq
The differential cross section 
which does not depend on the overall phase $\lambda$,
then reads
\beq
\label{X}
\frac{d\sigma}{d \cos \theta} = \frac{G_F^2}{2 \pi} \, X \quad ,
\eeq
with
\bea
\label{set2}
X &=& |a \, e^{i \d/2} \, f_1 + b \, e^{-i \d/2} \, f_2|^2 \quad \mbox{or}
\nonumber \\  
X &=& a^2 \, f_1^2 + 2 a b c \, f_1 f_2 + b^2 \, f_2^2 \quad , \nonumber \\  
f_i &=& \frac{tu-\frac{1}{2}(t+u)\, q_i}{(t- q_i)\,(u- q_i)} \quad (i=1,2)
\quad .
\eea
In eq. (\ref{set2}) $c=\cos \d$ is the cosine of the relative
phase between $m_{N_1} \,U_{eN_1}^2$ and   
$m_{N_2} \,U_{eN_2}^2$, while   
the quantities $t$, $u$, $s=t+u$ and $X=X ( t , u )$,
are understood to be measurable at any suitable 
finite set of values for the kinematic
variables. As seen from eq. (\ref{set2}), the reduced cross section $X$ 
depends on the 5 real parameters  
\beq
\label{res2}
\label{9}
a \, , \, b \, , \, q_1 \, , \, q_2 \,  \, > \, 0 
\hspace*{0.3cm} \mbox{and} \hspace*{0.3cm}
\ c = \cos \d
\hspace*{0.3cm} ; \hspace*{0.3cm}
-1 \le \ c \le 1
\eeq
The process \signal, in the sense
of a leading approximation, 
does not exhibit  direct CP (or T) violation. 
A nonvanishing CP violation only results, 
if in addition to the CP odd phase $\d$ also 
a CP even phase $\a$, generated by the electromagnetic interaction,
is involved \cite{PManant} 
(e.g. through Coulomb exchange in the initial state).
This is in analogy to direct CP violation in weak decays of
hadrons: $\d$ plays the role of a CKM angle, while
$\a$ is to be identified with a phase generated by the strong
interaction. 
In order to find CP violation experimentally, 
the pair of
CP associated reactions
\signal and \revsignal needs to be observed with high statistics,
an unrealistic endeavour for the two reactions at hand.
High statistics and precision are (would be) 
mandatory because it is the difference of 
respective differential 
cross sections which reveals the sought CP asymmetry.
This difference, which would be strictly zero 
neglecting electromagnetic
initial- or final state interaction,  
is much smaller (by a factor $\alpha$) than the
individual interference patterns. 

As the interference pattern  is much more
readily observable - due to its leading character - 
than any actually CP violating effect,
we concentrate on working out a strategy to etablish (or refute!) the
CP violating angle $\d$, which enters the differential cross
section through its cosine.
Assuming the discovery of the reaction \signal 
at some initial c.m. energy, predominantly due to
the exchange of a single (noninterfering) 
heavy Majorana neutrino, 
it is well conceivable that the sought interference 
can be established when going to higher energies.

From  eq. (\ref{set2}) 
it becomes clear that all the 5 parameters $a$, $b$, $c$, $q_1$, and
$q_2$ (within their respective ranges as shown in 
eq. (\ref{res2}))   
can be extracted in principle from suitable data,
assuming this interference scenario is realized in nature 
with a sufficiently large cross section.
To emphasize again, a value for $c=\cos \d$ different from 
$\pm 1$ would establish the existence of
CP violation.

As this issue might become relevant in future, 
we decided to elaborate a detailed stategy to extract
the values of the parameters from data. 
However, as the details of this discussion are necessarily
technical, we relegate it to the appendix. 

\section{Summary and Conclusions}
In this paper we have investigated what can be learnt
about the (heavy) neutrino sector from the reaction \signal
at future colliders. In a first scenario we have assumed
that only one of the heavy neutrinos is light enough to
lead to a detectable cross section. For this case we have
worked out discovery limits in the $(m_N, U_{eN}^2)$-plane.
We found that at the c.m. energy $\sqrt{s}=500$ GeV 
neutrinos with masses up to 2 TeV can be discovered, assuming
the relevant mixing angle to be at its present bound originating 
from low-energy experiments. At $\sqrt{s}=1$ TeV, the 
discovery region is much larger: Neutrinos with masses up to
12 TeV can be discovered.
For the discovery of heavy neutrinos
$e^-e^-$ colliders are better suited than $e^+e^-$ machines,
because in the latter case the neutrinos have to be produced;
consequently, one is only sensitive to neutrino masses  
smaller than $\sqrt{s}$. However, once a heavy neutrino is
found, its properties (like decay channels etc.) could be
better investigated in an $e^+e^-$ envirnment (with sufficient
energy). Therefore, as in many other aspects, $e^+e^-$ and 
$e^-e^-$ colliders are complementary.

Finally, we have investigated a scenario where two of the
heavy Majorana neutrinos are relatively light. 
There, in principle the relative phase $\delta$ between the
mixing parameters can be measured. This phase is responsible
for manifest CP violation in \signal: If 
$\d \neq 0,\pi$, direct CP violation is established.   
For this reason we have discussed a rather detailed
strategy how to extract this phase from data. 

\vspace*{1.5cm}
{\bf Acknowledgments:} We thank F. Cuypers, T. Han and C.A.
 Heusch for helpful discussions and 
the organizers of the Snowmass Workshop 
for a  wunderful time in Colorado.

\vspace*{1.5cm}
\centerline{\Large{\bf Appendix}}
\appendix
\section{Stategy to extract the CP violating phase}
As seen from eq. (\ref{set2}), the reduced cross section $X$
is a rational function in the variables $t$ and $u$, i.e.,
of the form $X = g/h$, where $g$ and $h$ are symmetric polynomials
in $t$ and $u$. For the following it is useful to write this
somewhat more explicitly as  
\beq
\label{Xpoly}
X=X(t,u) = \frac{g_{\m\n} \, t^\m u^\n}{h_{\a \b} \, t^\a u^\b} \quad , 
\eeq
where $g_{\m\n}=g_{\n\m}$ and $h_{\a \b}=h_{\b\a}$ are the coefficients 
of the symmetric polynomials
$g$ and $h$, respectively. 
In eq. (\ref{Xpoly}) the obvious summations
are tacitly understood. 
These coefficients themselves depend on the set of parameters
$\underline{\Pi}=[a, b, c, q_1, q_2]$ we finally want to extract,
more precisely
\beq
\label{19a}
h_{\alpha \beta} \ ( \ q_1 , q_2 \ )
\hspace*{0.3cm} , \hspace*{0.3cm}
g_{\mu \nu} \ ( \ a , b , c \ ; \ q_1 , q_2 \ )
\quad .
\eeq
For mathematical reason, as we will discuss later, it is
convenient to extract in a first step the coefficients
$g_{\m\nu}$ and $h_{\a\b}$ without making use of their
functional dependence on the parameters. Only after having determined
these coefficients we use the functional dependence
in order to extract the parameter set $\underline{\Pi}$. 
This has the most welcome advantage, that the analysis
done in this way 
becomes sensitive to scenarios where one heavy Majorana 
neutrino interferes with some "other mechanism" 
(with a dependence of the same form with respect to
 $t$ and $u$)
which also
contributes to the process \signal. As this process violates
lepton number conservation, this "other mechanism" necessarily
corresponds to new physics different from heavy Majorana
neutrino exchange.

\subsection{Extraction of the coefficients
 $g_{\m\n}$ and $h_{\a\b}$}
From the explicit form of the functions
$f_i$ in eq. (\ref{set2}), one easily finds that the indices (=powers
of $t$ and $u$)
$\m , \n , \a , \b$ in eq. (\ref{Xpoly}) range from 0 to 4.
In general such symmetric polynomials have 15
independent coefficients. 
More detailed inspection of the polynomial
$h$ shows that
the coefficient $h_{44}=1$, while the remaining coefficients of $h$
depend on the parameters $q_1$ and $q_2$. 
For the polynomial $g$ the coefficients
$g_{00}=0$, $g_{01} = 0$ (and therefore also $g_{10}=0$) while
the other 13 coefficients depend on the parameters 
$q_1$, $q_2$, $a$, $b$, and $c$. In total there are 14+13=27
nontrivial coefficients. 
We now put eq. (\ref{Xpoly}) into the form
\beq
\label{Xpolynew}
h_{\a\b} \,[t^\a  u^\b \, X] - g_{\m\n} \, t^\m  u^\n = 0 \quad .  
\eeq
In eq. (\ref{Xpolynew})
a linearly factorized form is achieved, separating the nonredundant
subset of determinable quantities
\beq
\label{18}
X_{\alpha \beta} = \frac{1}{2} \ ( \ t^\a u^\b 
+ t^\b u^\a \ ) \ X 
\hspace*{0.1cm} ; \hspace*{0.1cm}
T_{\mu \nu}  = \frac{1}{2} \ ( \ t^\m u^\n 
+ t^\n u^\m \ ) 
\eeq
from the set of quantities to be determined (=unknowns)
\beq
\label{19}
h_{\alpha \beta} 
\hspace*{0.3cm} , \hspace*{0.3cm}
g_{\mu \nu} 
\quad .
\eeq
Taking into account the constraints 
on the coefficients $g_{\m\n}$ and $h_{\a\b}$ 
mentioned in the text,
eq. (\ref{Xpolynew}) takes the inhomogeneous linear form
\beq
\label{20}
\begin{array}{l}
X_0 + X_{\alpha \beta} \ h_{\alpha \beta} 
+ T_{\mu \nu} \  (-g_{\mu \nu}) = 0
\hspace*{0.1cm} ( \rightarrow \Delta  \approx \ 0 ) 
\vspace*{0.1cm} \\
\begin{array}[t]{l}
\alpha \beta \neq 44
\vspace*{0.1cm} \\
\mu \nu \neq 00 , 10 , 01
\end{array}
\hspace*{0.3cm} ; \hspace*{0.3cm}
X_{0} = X_{44} = ( \, tu \, )^{4} \ X
\end{array}
\eeq
The expression $0 \  
( \rightarrow \Delta  \approx \ 0 )$ in eq. (\ref{20}) reflects
the fact that there are 
systematic deviations from the hypothetical relation in eq.
(\ref{20})  - theoretical as well as experimental - 
beyond the statistical errors, where the
latter are unproblematic from the mathematical point of view. 

We now form the two vectors of unknowns $z_{\ell}$ and 
associated determinables
$X_0,D_{\ell}$, restricting ourselves to the nonredundant set as specified 
in eq. (\ref{20})
\beq
\label{21}
\begin{array}{l}
\left \lbrace \, z_{\ell} \, \right \rbrace  = 
\left \lbrace \, h_{\alpha \beta} \, ; \, -g_{\mu \nu} \right \rbrace 
\hspace*{0.3cm} ; \hspace*{0.3cm}
\left \lbrace \, D_{\ell} \, \right \rbrace = 
\left \lbrace \, X_{\alpha \beta} \, ; \, T_{\mu \nu} \right \rbrace 
\vspace*{0.2cm} \\
\left \lbrace \, \ell = 1,...,14 \, \right \rbrace \leftrightarrow
\left \lbrace \, 0 \le \alpha \le \beta \le 4 \ ; \
\alpha \beta \neq 44 \, \right \rbrace 
\vspace*{0.1cm} \\
\left \lbrace \, \ell = 15,...,27 \, \right \rbrace \leftrightarrow
\left \lbrace \, 0 \le \mu \le \nu \le 4 \ ; \
\mu \nu \neq 00 \, , \, 01  \, \right \rbrace
\end{array}
\eeq
The association $ D_{\ell} \leftrightarrow 
\left \lbrace \, X_{\alpha \beta} \, ; \, T_{\mu \nu} \right \rbrace$
for $\alpha \neq \beta$ or $\mu \neq \nu$ is
\beq
\label{22}
\begin{array}{l}
D_{\ell} = 
\left \lbrace
\begin{array}{l}
2 X_{\alpha \beta} 
\hspace*{0.3cm} \mbox{for} \hspace*{0.3cm} 
\alpha <  \beta \leftrightarrow \ell = 1,...,14
\vspace*{0.1cm} \\
2 T_{\mu \nu} 
\hspace*{0.4cm} \mbox{for} \hspace*{0.3cm} 
\mu < \nu \leftrightarrow \ell = 15,...,27
\end{array}
\right .
\end{array}
\eeq
in order to take into account the symmetric summation relative to the indices
$\alpha \neq \beta$ and $\mu \neq \nu$. Then eq. 
(\ref{20}) is cast into conventional
linear form 
\beq
\label{23}
\begin{array}{l}
X_0 + \sum_{\ell = 1}^{27}
\ D_{\ell} \ z_{\ell}  = 0
\hspace*{0.1cm} ( \rightarrow \Delta  \approx \ 0 ) 
\vspace*{0.25cm} \\
\left \lbrace \, X_{0} \, , \, D_{\ell} \, , \, \Delta , \, \right \rbrace =
\left \lbrace \, X_{0} \, , \, D_{\ell} \, , \, \Delta , \, \right \rbrace
\, ( \, t \, , \, u \, )
\end{array}
\eeq
In eq. (\ref{23}) $ \left \lbrace \, X_{0} \, , 
\, D_{\ell} \, \right \rbrace $
are measured or exactly known functions, while $\Delta$ is an unknown 
function of the kinematic variables. The latter is set 0 whenever possible,
but should be kept in the equation in order to avoid eventual inconsistencies,
which typically arise when the signal (or the full details of the assumed
interference pattern) are insignificant relative to the systematically
misinterpreted background, systematic and/or statistical measurement errors.

To the nonredundant ${ \ell }$ set,
consisting of the functions and unknowns
$\left \lbrace \, D_{\ell} \, , \, z_{\ell} \, , \, X_{0} \ ; \ \Delta
\, \right \rbrace$ we associate - one by one - a given
$\left \lbrace k \right \rbrace$ set ($k = 1,...,27$) in the following way:
we {\bf choose} a set of weight functions over the region of experimentally
accessible kinematic variables $t,u$
\beq
\label{24}
\begin{array}{l}
\left \lbrace \ w_k \ \right \rbrace 
\hspace*{0.2cm} : \hspace*{0.2cm} 
w_{k} = w_{k} \ ( \, t \, , \, u \, )
\hspace*{0.2cm} ; \hspace*{0.2cm} 
k = 1,...,27
\vspace*{0.25cm} \\
\begin{array}{lll ll}
X_{0} & \rightarrow & \left ( \, X_{0} \, \right )_{\ k} & = &
{\displaystyle \int} \ dt du \ w_{k} \ X_{0}
\vspace*{0.2cm} \\
D_{\ell} & \rightarrow & D_{\ k \, \ell}  & = &
{\displaystyle \int} \ dt du  \ w_{k} \ D_{\ell}
\vspace*{0.2cm} \\
\Delta & \rightarrow & \Delta_{\ k}  & = &
{\displaystyle \int} \ dt du \ w_{k} \ \Delta
\end{array}
\end{array}
\eeq
The weighting functions $w$ include delta functions singling out 
\begin{description}
\item - individual
points or idealized bins $(t,u )_{k}$.
\item - line integrals, e.g. angular integrals with fixed c.m. energy.
\end{description}
There exists an infinite number of 
different $\left \lbrace k \right \rbrace$ sets,
each one defined through a given choice of 27 weighting functions 
$\left \lbrace  w_{k} \ \right \rbrace \ ; \ (k = 1,...27)$. 

We concentrate on a given set to be interpreted as sample case.
Eq. (\ref{23}) is thus transformed to an
inhomogeneous linear system of equations
\beq
\label{25}
X_0 \ + \ D \, z \ = \ 0 
\hspace*{0.1cm} ( \rightarrow \Delta  \approx \ 0 ) \quad , 
\eeq
where we use matrix notation in order to suppress
component indices.

In the present context the standard notions of linear algebra 
receive specific interpretations. The regular case applies
to a nonvanishing determinant of $D$.
Because of the approximate nature of eq. (\ref{25}), 
a vanishing and approximately vanishing determinant are equivalent
in a sense to be precisely defined.
The irregular case corresponds necessarily to the
situation where the data {\bf does not allow}
to derive the hypothetical interference pattern, i.e.
where the sought parameter $c = \cos \delta$ cannot be
significantly determined. 
While also this case constitutes useful information,
we concentrate now on the more interesting regular case.

In order to generate at least one $\left \lbrace k \right \rbrace$
set, at least 27 bins are needed. 
We choose one more 
in order to enable error analysis. 
This presupposes that at least
of order 140 to 240 events fall into these 28 bins.
As an example we may choose the 28 bins by taking 4 fixed c.m. energies 
and 7 angular bins for each; the actual choice, however, is irrelevant
for the following.   
One then can form
maximally 28 $\left \lbrace k \right \rbrace$ sets leaving out one of
the 28 bins in turn. We number each $\left \lbrace k \right \rbrace$ set
with the index $\left \lbrace n \right \rbrace \ $ ranging from 1 to 28.  

Thus 28 matrices $D^{\left \lbrace n \right \rbrace}$ result, which we
individually subject to predetermined criteria
\footnote{We do not discuss here details
of these criteria, straightforward to define,
which are to be used to eliminate the irregular case. } 
for qualifying as a regular matrix. 
These criteria will translate into
lower bounds for the determinants depending on 
$D^{\left \lbrace n \right \rbrace}$
\beq
\label{26}
\begin{array}{l}
\left | \, Det \ D^{\left \lbrace n \right \rbrace} \, \right |
\ > \ C \, ( \, \left \lbrace n \right \rbrace \, ) \ > \ 0
\quad .
\end{array}
\eeq
In general not all matrices $D^{\left \lbrace n \right \rbrace}$ will
pass the test in eq. (\ref{26}). Thus the last condition defining
the regular case is to specify the 
fraction of $\left \lbrace k \right \rbrace$ sets
for which the test in eq. (\ref{26}) is successful. 
To fix ideas we set this fraction to 75 \%.

Having finally defined the regular case, we 
assume in the following that the data qualifies
as regular. We retain only those  
$\left \lbrace k \right \rbrace$ sets, which pass the test in 
eq. (\ref{26}).  
We (re)number them with the label
$\left \lbrace n_{+} \right \rbrace$
\beq
\label{27}
\begin{array}{l}
n_{+} \ = \ 1,...,N_{+}
\hspace*{0.2cm} ; \hspace*{0.2cm}
21 \le N_{+} \le 28
\quad .
\end{array}
\eeq
The range for $N_{+}$, the number of regular $\left \lbrace k \right \rbrace$ 
sets, corresponds to the 75 \% limit required above.
We compare the calculational effort required to establish the
regular case with a direct $\chi^{2}$ minimalization procedure
to determine the five parameters at hand. Following the present
method the extraction of $28 \times 27 
\times 27 + 28  = 20'400$ quantities
 $X_{0 \, ,\, n}, D^{ \lbrace n \rbrace }$
was required. This corresponds to between 7 and 8 values for each
of the five parameters ( $7.3^{5} \approx 20'400$ ).

Now we determine the unknowns $z$ from eq. (\ref{25})
 setting the systematic
deviation vector $\Delta = 0$ for each retained 
$\left \lbrace k \right \rbrace$ set
\beq
\label{28}
\begin{array}{l}
\left ( \ z \ = \ - \ D^{-1} \ X_{0} \ \right )_{\ 
\left \lbrace n_{+} \right \rbrace}
\hspace*{0.2cm} ; \hspace*{0.2cm}
n_{+} \ = \ 1,...,N_{+}
\quad .
\end{array}
\eeq
We can assign weights to each regular 
$\left \lbrace k \right \rbrace$ set, in inverse proportion to 
the estimated quadratic standard deviation
\beq
\label{29}
\begin{array}{l}
p_{n} \ ; \ \sum p_{n} = 1
\quad .
\end{array}
\eeq
In eq. (\ref{29}) and in the following the index $+$ for
$n_{+} \ = \ 1,...,N_{+}$ is omitted.
We then form the mean value and relative error correlation matrix  
\beq
\label{30}
\begin{array}{l}
\overline{z} = \sum p_{n} \ z_{\left \lbrace n \right \rbrace} 
\hspace*{0.1cm} ; \hspace*{0.1cm}
\varrho^{2}_{\ell \ell'} = \sum p_{n} 
\ \begin{array}{c}
( \, z_{\left \lbrace n \right \rbrace} - \overline{z} \ )_{\, \ell} 
\ ( \, z_{\left \lbrace n \right \rbrace} - \overline{z} \ )_{\, \ell'}
\vspace*{0.2cm} \\
\hline  \vspace*{-0.3cm} \\
\overline{z}_{\, \ell} \ \overline{z}_{\, \ell'}
\end{array}
\end{array}
\eeq
The estimate of the error matrix $\varrho^{2}_{\ell \ell'}$ in eq. 
(\ref{30}) is only
sufficient to estimate the correlation matrix for $N_{+} - 1$ 
out of the $N=27$ components
$z_{\ell}$.
Nevertheless we can perform a $\chi^{2}$ minimization - 
not to be confused with
a statistical $\chi^{2}$ - 
on the systematic deviations $\Delta_{n}$ in eq. (\ref{25}),
which we have set to zero in the first approximation step displayed 
in eq. (\ref{28}).
To this end we cast the set of eqs. (\ref{28}) into the form
\beq
\label{31}
\begin{array}{l}
D^{\left \lbrace n \right \rbrace} \ \left 
( \ z - z_{\left \lbrace n \right \rbrace} \ \right ) 
\ = \ \Delta_{\left \lbrace n \right \rbrace} 
\hspace*{0.1cm} ; \hspace*{0.1cm}
z_{\left \lbrace n \right \rbrace} = - 
\ \left ( \, D^{\left \lbrace n \right \rbrace} \ \right )^{-1} 
\ X_{0}^{\left \lbrace n \right \rbrace}
\vspace*{0.2cm} \\
\chi^{2} = \sum p_n \ 
\ \Delta_{\left \lbrace n \right \rbrace}^{2} 
\vspace*{-0.2cm} 
\end{array}
\eeq
In eq. (\ref{31}) we have used the euclidean norm 
$\Delta_{\left \lbrace n \right \rbrace}^{2} = 
\sum_{k} ( \Delta_{\left \lbrace n \right \rbrace} )_{k}^{2}$ to
evaluate the systematic deviation 
of $\Delta_{\left \lbrace n \right \rbrace}$ from 0.

We determine the minimum and standard 
deviation of the function $\chi^{2} \, ( z )$
defined in eq. (\ref{31})
\beq
\label{32}
\begin{array}{l}
\chi^{2} - \chi^{2}_{min}
= \left ( \, z - \left \langle z \right \rangle \, \right )^{T}
\ \left \langle D^{2} \right \rangle 
\ \left ( \, z - \left \langle z \right \rangle \, \right ) 
\vspace*{0.2cm} \\
\chi^{2}_{min}  = \sum p_{n} \ \left \lbrack \, 
D^{\left \lbrace n \right \rbrace}
\ \left ( \, \left \langle z \right \rangle -
 z_{\left \lbrace n \right \rbrace}
\, \right )
\, \right \rbrack^{2}
\vspace*{0.2cm} \\
\left \langle z \right \rangle \ =
\ \left \langle D^{2} \right \rangle^{-1} 
\ \sum p_{n} \ D^{\left \lbrace n \right \rbrace \ T}
\, D^{\left \lbrace n \right \rbrace} \ z_{\left \lbrace n \right \rbrace}
\vspace*{0.2cm} \\
\left \langle D^{2} \right \rangle \ = 
\ \sum p_{n} \ D^{\left \lbrace n \right \rbrace \ T}
\, D^{\left \lbrace n \right \rbrace} 
\end{array}
\eeq
In eq. (\ref{32}) the symbol $T$ denotes transposition.

All quantities except the free variable $z$ are
functions of the determinables 
$ X_{0}^{\left \lbrace n \right \rbrace}
\ , \ D^{\left \lbrace n \right \rbrace}$,
depending in a highly nonlinear way on the matrix elements of  $D$. 
The regular case ensures that the (symmetric) matrix
$\left \langle D^{2} \right \rangle$ is regular. 

The $\chi^{2}$ procedure outlined here is equivalent to a
direct fit of the parameters obtained from the hypothesis in
the form of eq. (\ref{set2}), using the quadratic
deviation function
\beq
\label{33}
\chi'^{2} =
\left ( \, X \ - \  | 
a \, e^{i\d/2} \, f_1 + b \, e^{-i\d/2} \, f_2|^{\, 2}
\, \right )^{2}
\eeq
with one notable difference : the relation in eq. (\ref{32}),
quadratic in the free variable $z$,
is functionally exact. In the latter, hypothesis and
the determination of determinables are fully separated.
The functional dependence of the quantities
\beq
\label{34}
\begin{array}{l}
\left \lbrace \, \chi^{2}_{min} 
\, , \, 
\left \langle z \right \rangle
\, , \, 
\left \langle D^{2} \right \rangle
\, , \, 
z_{\left \lbrace n' \right \rbrace}
\, \right \rbrace
\ \left \lbrack X_{0 \ n}
\, , \, 
D^{\left \lbrace n \right \rbrace}
\, \right \rbrack
\end{array}
\eeq
on the arguments 
$X_{0 \ n} \, , \, D^{\left \lbrace n \right \rbrace}$ and $p_{n}$
is known and exact. The first two arguments,
defined in eqs. (\ref{18}) - (\ref{24}), involve, 
beyond the cross section,
known auxiliary functions used to formulate the hypothesis but as such
are independent of it. Their determination is subject
to systematic and statistical {\bf measurement errors only}.
These can be analyzed independently to any test of the
hypopthesis. 

Furthermore all errors, including the incomplete theoretical knowledge
of signal as well as remaining background amplitudes, are fully accounted
for by the unknown quantities $\Delta_{\left \lbrace n \right \rbrace}$
defining the function $\chi^{2}$ according to eq. (\ref{31}).

%

\underline{{\bf $\sigma$ contours}}
\vspace*{0.2cm}

The regular case ensures a pronounced minimum of 
$\Delta \chi^{2} \, ( \, z \, )$ at $z = \left \langle z \right \rangle$. 
This is not surprising, since the conditions in eq. (\ref{26}) defining
the regular case qualify the hypothesis as viable (necessary
but not sufficient conditions).

We consider the $(f-\sigma)$ contours ($f=1,2,...$) 
bounding the regions of $z$ values
\beq
\label{36}
\begin{array}{l}
\Delta \chi^{2} \, ( \, z \, ) = 
\left ( \, z - \left \langle z \right \rangle \, \right )^{T}
\ \left \langle D^{2} \right \rangle 
\ \left ( \, z - \left \langle z \right \rangle \, \right ) 
\ \le \ f^{2} \ \chi^{2}_{min}
\end{array}
\eeq
The probabilistic syntax inherent to the notion of $\sigma$ contour
is only true under the following circumstances 
\begin{description}
\item a) all systematic deviations are zero and the hypothesis is true.
\item b) the systematic deviations $\Delta_{\left \lbrace n \right \rbrace}$ 
are themselves independent stochastic variables and the hypothesis is true.
\end{description}

Case a) above applies in an approximative sense to the siuation
where statistical errors dominate. This will most probably be the case,
whenever the lepton violating signals discussed here will become observable
for the first time.

Nevertheless the value of $f$ to be applied 
to the following analysis is
a matter of initially judicious choice to be followed by subsequent
checks. The method at hand allows several such checks with the
same data. The first consists in comparing the value for
$\left \langle z \right \rangle$ with 
$\overline{z}$ defined in eq. (\ref{30})
\beq
\label{38}
\begin{array}{l}
\Delta \chi^{2} \, ( \, \overline{z} \, ) = \overline{f}^{2} \ \chi^{2}_{min}
\end{array}
\eeq
Irrelevant systematic deviations imply the relations involving
$\overline{f}$ in eq. (\ref{38}) 
\beq
\label{39}
\begin{array}{l}
\overline{f} \approx 1
\hspace*{0.2cm} ; \hspace*{0.2cm}
\overline{f}^{2} \ \chi^{2}_{min} = O \, 
\left ( \ \overline{z}^{T} \ \varrho^{2} \ \overline{z} 
\ || \ \left \langle D^{2} \right \rangle \ || \, \right )
\end{array}
\eeq
In eq. (\ref{39}) $\varrho^{2}$ denotes the variance matrix
estimated in eq. (\ref{30}) and $|| \ \left \langle D^{2} \right \rangle \ ||$
an appropriate norm, e.g 
$|| \ \left \langle D^{2} \right \rangle \ || = \frac{1}{N} \ tr 
\ \left \langle D^{2} \right \rangle$.

The tests outlined above can be  
extended, using the same data, choosing new bins,
thus enlarging the number of regular $\left \lbrace k \right \rbrace$ sets.
It is understood that throughout 
\beq
\label{40}
\begin{array}{l}
( \,  \left \langle z \right \rangle -  
z_{\left \lbrace n \right \rbrace} \, )^{2}
\, , \,  
( \, \overline{z} - z_{\left \lbrace n \right \rbrace} \, )^{2}
\, \ll \,
\left \langle z \right \rangle^{2} \, , \,  \overline{z}^{2}
\end{array}
\eeq
holds.
If no satisfactory factor $f = O \, (1)$ can be found, this
means that the hypothesis cannot be significantly established,
without seperate understanding of systematic deviations.

We continue the discussion on the contrary assumptions,
i.e. that a satisfactory ($f-\sigma$) contour is determined 
with $f \approx 1$ such that
\beq
\label{41}
\begin{array}{l}
\Delta \chi^{2} \, ( \, z_{true} \, ) 
\ \le \ f^{2} \ \chi^{2}_{min} 
\ \rightarrow \ z_{true} \ \approx \ \left \langle z \right \rangle 
\end{array}
\eeq

\subsection{Establishing the interference scenario }
\vspace*{0.2cm}
All disqualifying criteria being positively met, still does not prove the
hypothesis of the interference scenario defined in eqs. (\ref{scenario2})
- (\ref{set2}), because so far we have only determined successfully
the vector $z$, which is nothing but the coefficients $g_{\m\n}$
and $h_{\a\b}$ in eqs. (\ref{19a}) and (\ref{19}) [see also eq. (\ref{21})].
The vector $z$ depends on the set of parameters
$\underline{\Pi}=[q_1,q_2,a,b,c]$, in which we are ultimately
interested in. 
For each point in the parameter set $\underline{\Pi}$ 
we calculate $z=z(\underline{\Pi})$. The so determined vectors 
$z$
sweep out a 5 dimensional subspace, denoted by $S_{5}$,
when varying the five parameters $\underline{\Pi}$ in the allowed 
range specified in eq. (\ref{9}).

The ($f - \sigma$) allowed region in the parameter 
space $\underline{\Pi}$ 
is the intersection of $S_{5}$ with the $z$ region bounded
by the corresponding contour according to eq. (\ref{36})
\beq
\label{44}
\begin{array}{l}
\left \lbrace
\ f - \sigma \ \left \lbrack \ \underline{\Pi} \ \right \rbrack
\ \right \rbrace
\ = \ \left \lbrace 
\ \left . \underline{\Pi} \ \right |
\ \Delta \chi^{2} 
\ \left ( \ z \ \left \lbrack \ \underline{\Pi} 
\ \right \rbrack \ \right ) 
\ \le \ f^{2} \ \chi^{2}_{min} \ \right \rbrace
\vspace*{0.2cm} 
\end{array}
\eeq
The final test of our hypothesis of an interference scenario
of two heavy neutrino flavors, requires the region of
allowed parameters 
($f - \sigma \ \left \lbrack \ \underline{\Pi} \ \right \rbrack$)
not to be empty or concentrated very near to the bounding 
($f - \sigma$) contour. 
The best values for the five parameters 
$\left \lbrack \ \underline{\Pi} \ \right \rbrack$ are then
determined by the minimum of
the function
$\Delta \chi^{2} \ \left ( \ z \ \left \lbrack \ \underline{\Pi} 
\ \right \rbrack \ \right )$ in eq. (\ref{44}) 
\beq
\label{45}
\begin{array}{l}
\Delta \chi^{2} \ \left ( \ z \ \left \lbrack \ \underline{\Pi}_{best} 
\ \right \rbrack \ \right )
\ = \ \mbox{Min}_{\ \left \lbrack \, \underline{\Pi} \, \right \rbrack}
\ \Delta \chi^{2} 
\ \left ( \ z \ \left \lbrack \ \underline{\Pi} 
\ \right \rbrack \ \right ) 
\ < \ f^{2} \ \chi^{2}_{min} 
\vspace*{0.2cm} 
\end{array}
\eeq
Amusingly, an interesting situation also arises 
if this test fails, whereby the specific
hypothesis (i.e., the interference of 2 Majorana neutrinos) 
is definitely disproved. However, on the level of polynomials of the
same degree as those defining the hypothesis, an alternative interference 
pattern is established by the consistent form of the 
($f - \sigma$) contour defined in eq. (\ref{41}).
This positively would prove the interference of the one flavor signal,
considered as established according to the criteria discussed in the
previous sections, albeit with an amplitude of another nature
than the one considered, also violating
lepton number! Interference with a standard model background amplitude
is by definition excluded.

We continue the discussion for the alternative case where
the criteria 
in eqs. (\ref{44}) and (\ref{45}) are satisfied.
The full amplitude for two interfering
Majorana neutrinos 
can be reconstructed for all values of the parameters 
\beq
\label{46}
\begin{array}{l}
q_1 = \left ( m_{N_1} \right )^{2}
\ ; \ q_2 = \left ( m_{N_2} \right )^{2}
\ ; \ q_1 < q_2  
\ , \ a , b \ge 0 
\vspace*{0.2cm} \\
\ c 
\hspace*{0.3cm} ; \hspace*{0.3cm}
-1 \le \ c \le 1
\end{array}
\eeq
in the allowed region $\left \lbrace
\ f - \sigma \ \left \lbrack \ \underline{\Pi} \ 
\right \rbrack \ \right \rbrace$.

On this basis additional tests of the consistency of parametrization with
the data can be performed. 
Furthermore the absolute values of the mixing
parameters $| \ U_{eN_1} \ |$ and $| \ U_{eN_2} \ |$ in eq. (\ref{7}) 
can be determined.
These are subject to the constraints unrelated to the
data sample considered here \cite{signal}
\beq
\label{47}
\begin{array}{l}
| \ U_{eN_1} \ |^{2} \ + \ | \ U_{eN_2} \ |^{2} \ \le \ 4 \times 10^{-4}
\end{array} \quad ,
\eeq
eventually narrowing down the acceptable parameter region.

Finally we turn to the most subtle of the five parameters,
$c$, the cosine of the relative phase between
$m_{N_1} \, U_{eN_1}^{2}$ and 
$m_{N_2} \, U_{eN_2}^{2}$. The significant determination of $c$
relies on a solid error analysis with respect to the other 4 parameters,
since the error matrix must show a high degree of correlation between
the errors of $c$ and those of the other parameters. In the vein of the 
present discussion we assume all 
further tests to reveal no inconsistencies.
Then we foresee the result
\beq
\label{48}
\begin{array}{l}
-1 \ \le c_{-} \ \le \ c_{best} \ \le \ c_{+} \ \le \ 1
\end{array}
\eeq
according to the ($f - \sigma$) criteria defined in eqs. 
(\ref{41}) , (\ref{44}) and (\ref{45}).

Three distinct possibilities arise, in ascending order of interest:
\begin{description}
\item 1) 
\hspace*{0.1cm} 
$c_{+}$ is not significantly different - within $(2 - \sigma)$ say - 
from 1, {\bf nor} is $c_{-}$ from -1.
\item 2) 
\hspace*{0.1cm} 
{\bf either} $c_{+}$ is not significantly different  
from 1, {\bf or} $c_{-}$ insignificantly from -1.
\item 3) 
\hspace*{0.1cm} 
$c_{+}$ {\bf is} significantly different  
from 1, {\bf and} $c_{-}$ from -1.
\end{description}

Case 1) is unlikely, given the severe criteria in particular those
regarding the regular case, discussed above, but cannot be
entirely excluded. The conclusion is that the interference
pattern can be established for some part of the allowed parameter
region but the error margin is too large for an actual 
determination of $c$.

Case 2) establishes the interference pattern but does not exclude
the CP conserving relative phases $\delta = 0 \ , \ \pi$.

Finally case 3), which serves as motivation of this analysis,
not only establishes - within the errors - the interference pattern but proves,
albeit indirectly, the CP violating character of
this interference. 

We shall conclude this discussion on the 
hypothesis that this last case prevails.

%

\end{document}